\documentstyle[draft]{mn}

\newif\ifAMStwofonts

\input psfig.sty

\ifoldfss
  \ifCUPmtlplainloaded \else
    \NewTextAlphabet{textbfit} {cmbxti10} {}
    \NewTextAlphabet{textbfss} {cmssbx10} {}
    \NewMathAlphabet{mathbfit} {cmbxti10} {} 
    \NewMathAlphabet{mathbfss} {cmssbx10} {} 
  \fi
  \ifAMStwofonts
    \ifCUPmtlplainloaded \else
      \NewSymbolFont{upmath} {eurm10}
      \NewSymbolFont{AMSa} {msam10}
      \NewMathSymbol{\upi}     {0}{upmath}{19}
      \NewMathSymbol{\umu}     {0}{upmath}{16}
      \NewMathSymbol{\upartial}{0}{upmath}{40}
      \NewMathSymbol{\leqslant}{3}{AMSa}{36}
      \NewMathSymbol{\geqslant}{3}{AMSa}{3E}

       \let\le=\leqslant
       
    \fi
  \fi
\fi 

\ifnfssone
  \newmathalphabet{\mathit}
  \addtoversion{normal}{\mathit}{cmr}{m}{it}
  \addtoversion{bold}{\mathit}{cmr}{bx}{it}
  \newmathalphabet{\mathbfit} 
  \addtoversion{normal}{\mathbfit}{cmr}{bx}{it}
  \addtoversion{bold}{\mathbfit}{cmr}{bx}{it}
  \newmathalphabet{\mathbfss} 
  \addtoversion{normal}{\mathbfss}{cmss}{bx}{n}
  \addtoversion{bold}{\mathbfss}{cmss}{bx}{n}
  \ifAMStwofonts
    \ifCUPmtlplainloaded \else
      \UseAMStwoboldmath
      \makeatletter
      \new@mathgroup\upmath@group
      \define@mathgroup\mv@normal\upmath@group{eur}{m}{n}
      \define@mathgroup\mv@bold\upmath@group{eur}{b}{n}
      \edef\UPM{\hexnumber\upmath@group}
      \new@mathgroup\amsa@group
      \define@mathgroup\mv@normal\amsa@group{msa}{m}{n}
      \define@mathgroup\mv@bold\amsa@group{msa}{m}{n}
      \edef\AMSa{\hexnumber\amsa@group}
      \makeatother
      \mathchardef\upi="0\UPM19
      \mathchardef\umu="0\UPM16
      \mathchardef\upartial="0\UPM40
      \mathchardef\leqslant="3\AMSa36
      \mathchardef\geqslant="3\AMSa3E

       \let\le=\leqslant

    \fi
  \fi
\fi 

\ifnfsstwo
  \DeclareMathAlphabet{\mathbfit}{OT1}{cmr}{bx}{it}
  \SetMathAlphabet\mathbfit{bold}{OT1}{cmr}{bx}{it}
  \DeclareMathAlphabet{\mathbfss}{OT1}{cmss}{bx}{n}
  \SetMathAlphabet\mathbfss{bold}{OT1}{cmss}{bx}{n}
  \ifAMStwofonts
    \ifCUPmtlplainloaded \else
      \DeclareSymbolFont{UPM}{U}{eur}{m}{n}
      \SetSymbolFont{UPM}{bold}{U}{eur}{b}{n}
      \DeclareSymbolFont{AMSa}{U}{msa}{m}{n}
      \DeclareMathSymbol{\upi}{0}{UPM}{"19}
      \DeclareMathSymbol{\umu}{0}{UPM}{"16}
      \DeclareMathSymbol{\upartial}{0}{UPM}{"40}
      \DeclareMathSymbol{\leqslant}{3}{AMSa}{"36}
      \DeclareMathSymbol{\geqslant}{3}{AMSa}{"3E}

       \let\le=\leqslant

    \fi
  \fi
\fi 

\ifCUPmtlplainloaded \else
  \ifAMStwofonts \else 
    \def\upi{\pi}
    \def\umu{\mu}
    \def\upartial{\partial}
  \fi
\fi

\title{Nuclear star formation in the hot-spot galaxy NGC~2903}
\author[A. Alonso-Herrero et al.]
       {A. Alonso-Herrero$^1$\footnote{email: aalonso@star.herts.ac.uk}, 
        S. D. Ryder$^2$ and J. H. Knapen$^{1,3}$\thanks{Visiting
Astronomer, Canada--France--Hawaii Telescope operated by the National
Research
Council of Canada, the Centre National de la Recherche Scientifique de
France and the University of Hawaii.}\\
        $^1$University of Hertfordshire, Department 
of Physical Sciences, College Lane, Hatfield, Herts AL10 9AB, UK\\
        $^2$Anglo-Australian Observatory, P. O. Box 296, 
        Epping, NSW 1710, Australia\\
        $^3$Isaac Newton Group of Telescopes, Apartado 321, 
        E-35700 Santa Cruz de La Palma, Spain}
\date{Accepted .
      Received ;
      in original form }

\pagerange{\pageref{firstpage}--\pageref{lastpage}}
\pubyear{1994}

\begin{document}

\maketitle

\label{firstpage}

\begin{abstract}

We present high-resolution near-infrared imaging 
obtained using adaptive optics 
and {\it HST}/NICMOS, and ground-based spectroscopy of the hot-spot galaxy 
NGC~2903. Our near-infrared resolution 
imaging enables us to resolve the infrared hot spots into individual 
young stellar clusters or groups of these. The spatial distribution of the 
stellar clusters is not coincident with that of the bright 
H\,{\sc ii} regions, as revealed by the {\it HST}/NICMOS Pa$\alpha$
image. Overall, the circumnuclear star formation in NGC~2903 shows a 
ring-like morphology with an approximate diameter of 625\,pc. 

The SF properties of the stellar clusters and H\,{\sc ii} regions 
have been studied using the photometric and spectroscopic 
information in conjunction with evolutionary synthesis models. 
The population of bright stellar clusters shows a very narrow
range of ages, $4-7 \times 10^6\,$yr after the peak of star formation,
or absolute ages $6.5-9.5 \times 10^6\,$yr 
(for the assumed short-duration Gaussian bursts), and luminosities
similar to the clusters found in the Antennae interacting galaxy. This
population of young stellar clusters accounts for some $7-12\%$ of 
the total stellar mass in the central 625\,pc of NGC~2903. 
The H\,{\sc ii} regions in the ring 
of star formation have luminosities close to that of the 
super-giant H\,{\sc ii} region 30 Doradus, they are younger than 
the stellar clusters, and will probably evolve into  bright 
infrared stellar clusters similar to those observed today. 
We find that the star formation 
efficiency in the central regions of NGC~2903 is higher than in 
normal galaxies,  approaching 
the lower end of infrared luminous galaxies. 

\end{abstract}

\begin{keywords}
galaxies: individual: NGC~2903 -- infrared: galaxies -- galaxies: photometry
-- galaxies: star clusters -- galaxies: starburst
\end{keywords}

\section{Introduction}

Numerical simulations of  bars have shown that 
they may provide an efficient mechanism for transporting interstellar 
gas into the central regions of galaxies. Thus, 
bars are often invoked as a possible 
mechanism to fuel nuclear/circumnuclear
star formation (SF; Heller \& Shlosman 1994), as well as 
active galactic nuclei (Shlosman, Frank, \& Begelman 1989). 
It is now well established that circumnuclear regions 
(CNRs) of enhanced SF are commonly associated with the
presence of stellar bars (see the recent review by Knapen 1999, and
references therein). In particular, a significant percentage of barred 
galaxies show nuclear rings within 1\,kpc from the nucleus of
the galaxy; these are believed to be 
formed by gas accumulation near galactic resonances. 
In these nuclear rings, strong density 
enhancements of gas and increased SF are often observed  
(Piner, Stone, \& Teuben 1995; Knapen et al. 1995b; Buta \&
Combes 1996; Shlosman 1999).

Knapen (1999) pointed out that CNRs are excellent laboratories 
for the study of the effects of inflow processes on the triggering 
of SF. A detailed study of CNRs has been hampered by the lack of 
imaging at sufficiently high spatial resolution. The main  
difficulty arises when trying to 
disentangle the effects of young and old stars, cold gas and dust, and 
possibly emission from dust heated by the SF activity (e.g. Knapen et al. 
1995a, b). High resolution imaging allows one 
to resolve individual SF sites, 
to determine their location with respect to dust lanes, and to 
discover dynamically important morphological features that remain hidden in 
imaging of CNRs at lower resolution. These features include nested bars 
(e.g. Knapen et al. 1995a; Phillips et al 1996; Elmegreen et al. 
1998; Carollo, Stiavelli, \& Mack 1998; Colina \& Wada 2000), leading spiral arms 
(Knapen et al. 1995a, 2000); nuclear flocculent 
(e.g., Elmegreen et al. 1999) or even grand-design spirals (Laine 
et al. 1999).

High resolution near-infrared  (NIR) observations of CNR are 
of particular importance 
because they are less affected by extinction, and
because the NIR emission is a better tracer of old stellar 
population (and hence of the stellar mass distribution) 
than optical or ultraviolet observations. 
However, in sites of strong SF there can be 
a significant contribution from young red supergiants   
to the NIR emission, and therefore 
both NIR spectroscopy and imaging are an absolute need to study the 
SF processes in CNR  (e.g., Engelbracht et al. 1998; Ryder \& Knapen 1999; 
Puxley \& Brand 1999; Alonso-Herrero et al. 2000; 2001). 
Quantifying the contribution of young stars to the
NIR emission so the emission from old stars can be isolated is essential 
for the interpretation of the dynamics 
of CNRs and of inflow in barred galaxies in general, through the use of 
dynamical modeling. Such modeling is generally based upon an 
estimate of the 
galactic potential as derived from red or NIR imaging, under the assumption 
of a constant mass-to-light ratio (e.g., Quillen, Frogel, \&
Gonz\'alez 1994; Knapen et al. 1995a). It is important that this assumption be 
tested, and the circumstances under which the mass-to-light 
ratio may vary be understood, to allow for more accurate 
modelling.

In this paper we present high angular resolution  NIR adaptive optics 
(AO) and {\it HST}/NICMOS imaging, and  spectroscopy of 
the hot-spot galaxy NGC~2903. Although NGC~2903 is classified 
as an SAB(rs)bc (de Vaucouleurs et al. 1991), the presence of a 
large scale bar with a position angle (P.A.) of $\simeq 20\deg$ \ is clearly
inferred from $K$-band (Regan \& Elmegreen 1997) and 
H$\alpha$ (Jackson et al. 
1991; P\'erez-Ram\'{\i}rez \& Knapen 2000) images, as well as from CO 
observations (Regan, Sheth, \& Vogel 1999). Early NIR  studies of this 
galaxy have shown that considerable SF is occurring within
the complex hot spot morphology in the nuclear region 
(e.g., Wynn-Williams \& Becklin 1985; 
Simons et al. 1988). However these studies lacked the
necessary spatial resolution to determine whether 
the SF was confined to these hot spots 
or not. The infrared luminosity ($8-1000\,\mu$m) of NGC~2903 
is $9.1 \times 10^9\,{\rm L}_\odot$ (Sakamoto et al. 1999) which is 
similar to (although slightly smaller than) those of M82 and NGC~253, 
both also well known hot-spot galaxies. Throughout the paper we will 
assume a distance to NGC~2903 of $d = 8.6\,$Mpc (Telesco \& Harper 1980),
for which $1\,{\rm arcsec} \simeq 42\,$pc. This paper is organized as follows. 
Section~2 describes the observations. Section~3
discusses the morphology, and Section~4 the extinction. The properties of 
the stellar clusters and H\,{\sc ii} regions are 
studied in Sections~5 and 6, while conclusions are given in Section~7.

\section{Observations}

\subsection{CFHT AO $K^\prime$-band imaging}

We obtained a $K^\prime$-band image of the central region of 
NGC~2903 on 1998 October 25 with the Canada France Hawaii 
Telescope (CFHT), equipped with the adaptive optics (AO) system PUEO 
(Rigaut et al. 1999) and the KIR camera,
a high resolution $1024 \times 1024$ NIR camera based on a Rockwell
HAWAII HgCdTe array.  PUEO uses a curvature mirror and a wavefront
sensor to give, under favourable circumstances, diffraction limited
images at $H$ and $K$.  We used the nucleus of the galaxy itself as the
AO guide source, which is not a point source and thus does not allow
PUEO to reach this full correction. We estimate that the natural seeing
of around 0.6 arcsec \ at the time of the observations was improved to
to a resolution in the final image of about 0.25 arcsec.  The total 
integration time was 240 seconds, made up
from 4 individual dithered exposures.  The individual images were
sky-subtracted, flatfielded, corrected for bad pixels and combined into
the final mosaic. KIR's image scale is $0.03438 
\pm0.0007\,$arcsec \ pixel$^{-1}$,
and the total field of view of the individual exposures is
$36\,{\rm arcsec} \times 36\,{\rm arcsec}$.  
As the nucleus of the galaxy was positioned at
various locations across the array during individual exposures, the
total field of view of the final images is slightly larger 
($46\,{\rm arcsec} \times 46\,{\rm arcsec}$). The fully reduced 
image is presented in  the left panel of Figure~1.

Flux calibration of the image was performed by observing
standard stars from the list of Hawarden et al. (2000). The uncertainty 
in the photometric calibration of the image is about 
0.05\,mag. 

\subsection{{\it HST}/NICMOS observations}

We obtained NIR observations of NGC~2903 from the {\it HST} archive.  
The images were taken with camera NIC2  
(pixel size 0.076\,arcsec \ pixel$^{-1}$) 
through the F160W filter
(equivalent to a ground-based $H$-band filter), 
and with camera NIC3  (pixel size 0.2\,arcsec \ pixel$^{-1}$) taken 
through filters F160W and F187N (the latter filter contains
the emission line Pa$\alpha$ and continuum at $1.87\,\mu$m). The NIC3
observations are part of the survey of nearby galaxies 
conducted by B\"oker et al. (1999). 

Standard data reduction procedures were applied (see Alonso-Herrero
et al. 2000 for more details). The reduced NICMOS images were rotated to the 
usual orientation (north up, east to the left).
The spatial resolutions (FWHM) of the NIC2 and NIC3 images 
are 0.15\,arcsec (6\,pc) \ and 0.3\,arcsec, respectively, as measured from 
unresolved sources in the images. These sources show diffraction rings so 
we can be certain that good estimates of the resolution can be obtained from 
them. The flux calibration was performed using
the conversion factors based on measurements of the standard star P330-E
taken 
during the Servicing Mission Observatory Verification (SMOV) programme 
(M.J. Rieke 1999, private communication). Unfortunately there are 
no observations available of the adjacent continuum  to Pa$\alpha$, so 
we used the flux calibrated continuum image at $1.6\,\mu$m. 
Ideally continuum images on both 
sides of the emission line are needed to perform an accurate continuum 
subtraction which takes into account the spatial distribution of 
the extinction. However, the continuum subtracted 
Pa$\alpha$ image (flux calibrated) does not show negative values 
suggesting that the continuum subtraction was relatively accurate and 
that the differential extinction between 
$1.60\,\mu$m and $1.87\,\mu$m does not play a dominant role.
The continuum subtracted Pa$\alpha$ image is presented in the 
right panel of Figure~1.

An $H-K^\prime$ colour map was constructed using the {\it HST}/NICMOS 
$H$-band and the AO
$K^\prime$-band images. The process involved rebinning of the 
$K^\prime$-band image to the 
same pixel size as the NICMOS image and smoothing of the NICMOS image
to match the effective resolution  of the AO image. The 
$H-K^\prime$ colour map along with the NIC2 F160W image are presented
in Figure~2.

\begin{figure*}
\hspace{-15cm}
\psfig{file=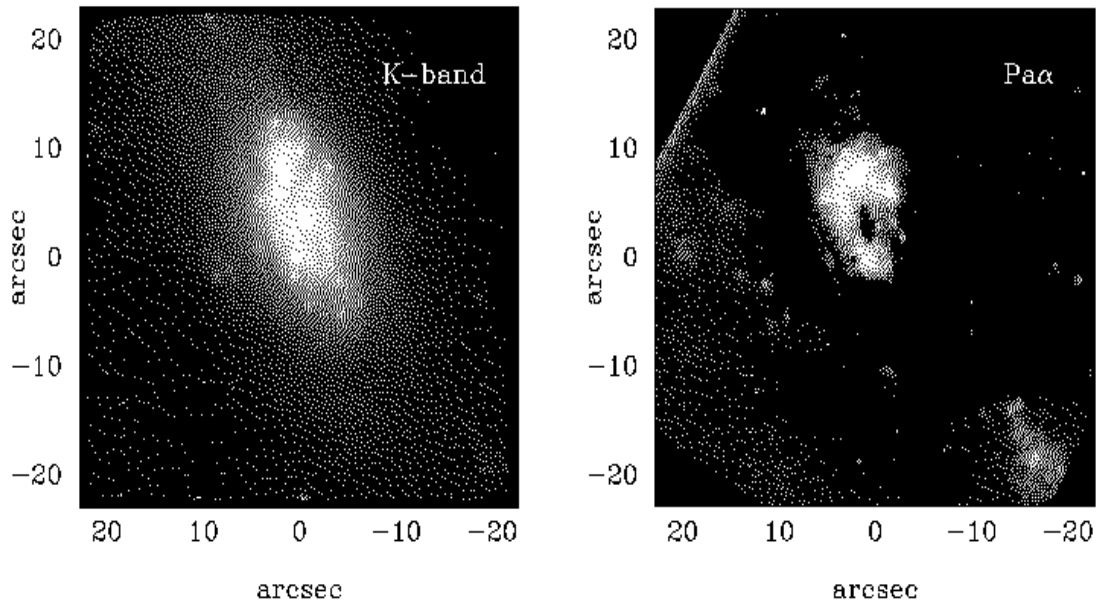,height=14cm,width=16cm,rheight=14cm,rwidth=-1cm,angle=0}
\vspace{-2cm}
\caption{{\it Left panel:} CFHT/PUEO AO $K^\prime$-band image. 
{\it Right panel:}   
Continuum-subtracted Pa$\alpha$ (NIC3 F187N) image. Both images 
are displayed
on a logarithmic scale. The  field of view is 
$46\,{\rm arcsec} \times 46\,{\rm arcsec}$
and the orientation is north up, east to the left.}
\end{figure*}

\begin{figure*}
\hspace{-15cm}
\psfig{file=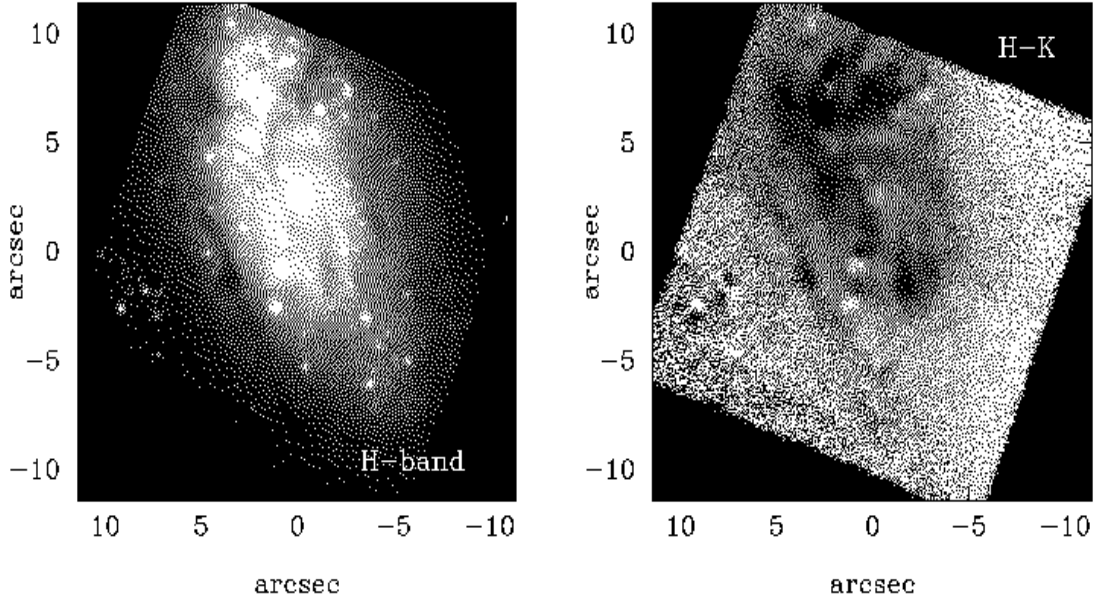,height=14cm,width=16cm,rheight=14cm,rwidth=0cm,angle=0}
\vspace{-2cm}
\caption{{\it Left panel:} $H$-band (NIC2 F160W) image displayed 
on a logarithmic scale. 
{\it Right panel}: $m_{\rm F160W}-K^\prime$  (equivalent to 
a ground-based $H-K^\prime$) colour map. 
Dark colours indicate regions with 
high extinction. The  $m_{\rm F160W}-K^\prime$ 
colour map is displayed from 0.1 to 0.8\,mag. 
The field of view of both images is $22\,{\rm arcsec} \times 
22\,{\rm arcsec}$.}
\end{figure*}

\begin{figure}
\psfig{file=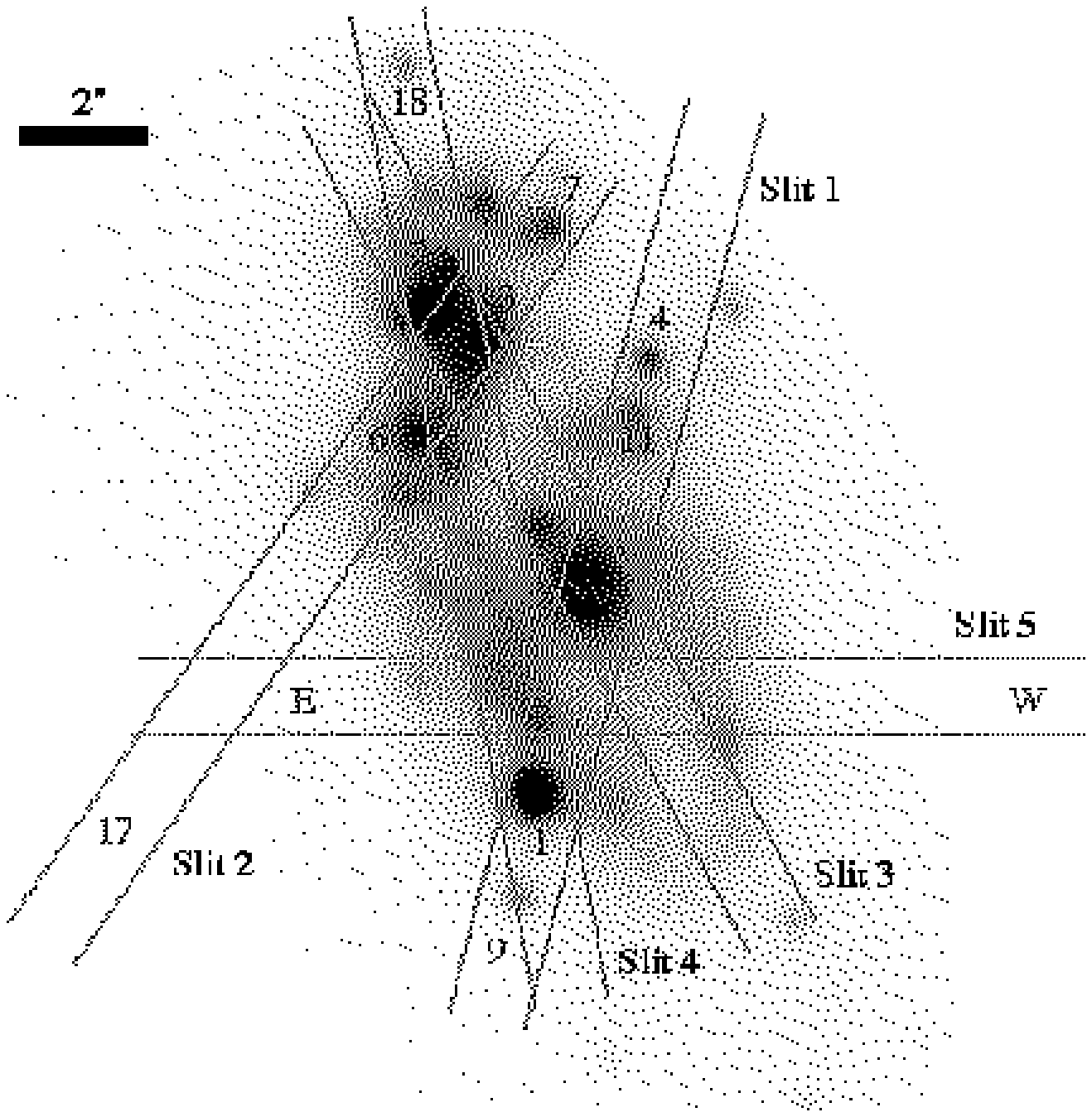,height=10cm,width=8cm,rheight=10cm,rwidth=8cm,angle=0}
\vspace{-1cm}
\caption{Location of the CGS4 slit positions observed,
together with identifications of the stellar clusters
extracted, marked on the AO $K^\prime$-band image.
The numbers refer to the cluster ID in Tables~2 and 3,
except for clusters 17 and 18 (which are outside
the field of the {\it HST}/NICMOS image), and the positions
marked "E" and "W" along Slit 5, which do not correspond
to any particular cluster but instead sample the
background stellar continuum.}
\end{figure}

\subsection{NIR Spectroscopy} 

Long-slit spectroscopy of the central regions of NGC~2903
was carried out on 1998 December 25-26 UT, and on 1999
April 1 UT, using the common-user NIR spectrograph
CGS4 (Mountain et al. 1990) on the United Kingdom Infrared
Telescope (UKIRT) at Mauna Kea. The observing setup was
essentially identical to that of Ryder, Knapen \& Takamiya
(2000), viz. a 1.2\,arcsec \ slit, 0.61\,arcsec pixel$^{-1}$, 
and a 40\,line mm$^{-1}$ grating in first order,
yielding complete spectral coverage between 1.85 and 
$2.45\,\mu$m at an effective resolving power $R\sim450$. Using
the AO $K^\prime$-image (Figure~1, left panel), a total of five slit
position angles and nuclear offsets were selected, as
marked on Figure~3 (see also Table~1); four of these cover at 
least 3 stellar clusters each, while the fifth samples the
background stellar continuum.

After setting the image rotator to the desired slit
position angle, a manual search for the $K$-band nuclear
peak was conducted, following which accurate offsets were
applied by use of the UKIRT crosshead to place the slit
at exactly the positions marked on Figure~3. During observations,
the detector array was micro-stepped by half a resolution
element every 30 seconds, to sub-sample the spectra and to
compensate for bad pixels; after four such steps, the telescope
itself was nodded by 36.6 arcseconds along the slit to collect
the sky spectrum, while recording the object spectra on the
opposite half of the detector. These object-sky exposure pairs
were differenced, and the result co-added until the total on-source
integration times listed in Table~1 were achieved. All observations
of NGC~2903 were bracketed by observations of the A2 V star BS 3657
at a similar airmass, to assist with removal of some of the telluric
features.

Data reduction has been carried out using a combination of
tasks within the Starlink CGS4DR and NOAO {\sc iraf}\footnote{
{\sc iraf} is distributed by the National Optical Astronomy
Observatories, which are operated by the Association
of Universities for Research in Astronomy, Inc., under
cooperative agreement with the National Science Foundation.}
packages. After flatfielding of the co-added object-sky pairs,
any residual sky lines were removed by fitting a low-order
polynomial to each row of the spatial axis. Plots of the
variation of continuum emission along the slit were compared
with the AO $K^\prime$-band image to define extraction apertures
appropriate to each cluster. In several cases it proved impossible, 
at the 0.61\,arcsec \ pixel$^{-1}$ scale of CGS4, to
separate the spectra of neighboring clusters, so the net 
emission from both was extracted. After extraction, the spectra
were wavelength-calibrated using observations of an Argon
arc lamp. The strong Br$\gamma$ absorption intrinsic to
BS 3657 was interpolated over, prior to dividing its spectrum
into that of each of the clusters, and scaling by a blackbody
curve corresponding to the temperature (8810\,K) and flux density
($K=6.46$) of BS 3657. Figure~4 shows the fully reduced spectra 
of clusters 2+3 (in slit 4) and the W position (in slit 5).

\begin{figure}
\psfig{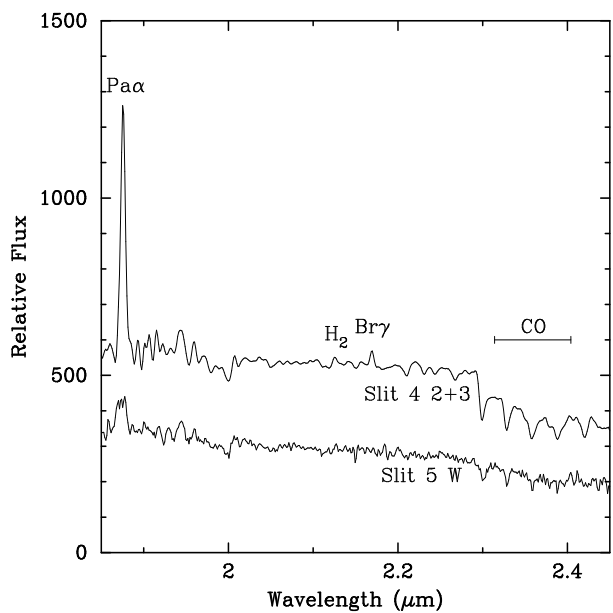}

\caption{NIR spectra of clusters 2 + 3 from Slit 4
(top), and background stellar continuum from W position of
Slit 5 (see Figure~3 and Table~1). The locations of the prominent
emission features, as well as the extent of the CO band, are
marked. The features between 1.9 and $2.0\,\mu$m are nearly all
telluric and standard star residuals.}
\end{figure}

  To define an objective continuum level right across the region
of extensive CO absorption redward of 2.3 $\mu$m, we fitted a 
power-law of the form $F_{\lambda} \propto \lambda^{\beta}$ to
featureless sections of the cluster spectra near 2.11 and 
$2.27\,\mu$m (rest wavelength), and normalised by this fit. The
equivalent width (EW) of the Br$\gamma$ emission line was determined
by fitting a Gaussian to the profile using the {\it splot} task in
{\sc iraf}. The CO spectroscopic index, as defined by Doyon, Joseph,
\& Wright (1994) has also been measured:

\begin{equation}
{\rm CO}_{\rm sp} = -2.5\log \langle R_{2.36} \rangle
\end{equation}

\noindent where $R_{2.36}$ is the mean normalised intensity between 2.31
and $2.40\,\mu$m in the rest frame of the galaxy (for NGC 2903,
this corresponds to the wavelength interval $2.314-2.404\,\mu$m).
At the (comparatively low) resolution of these observations, this
measure of the CO absorption due to young supergiants was found
by Ryder et al. (2000) to be more reliable than the narrower
interval favoured by Puxley, Doyon, \& Ward (1997). The equivalent
photometric CO index has then been computed using the transformation
given by Doyon et al. (1994):

\begin{equation}
{\rm CO}_{\rm ph} = ({\rm CO}_{\rm sp} + 0.02)/1.46
\end{equation}

\noindent for direct comparison with the evolutionary synthesis 
models (Section~5.2). In Table~2 we 
give the EW of Br$\gamma$ and photometric and spectroscopic 
values of the CO index and their corresponding errors sorted by 
the slit number. In cases where the same cluster(s) were observed 
by different slits, we find extremely good agreements (e.g., 
the nucleus in slits 1 and 3,
and clusters 2+3 in slits 3 and 4). The one exception is clusters 1+9
in slits 1 and 4, but as can be seen from Figure~3, there is less flux
from cluster 9 in slit 4. The slit 5 regions E and W
(7.3\,arcsec \ E and W of the nucleus
as shown in Figure~3) have only upper limits on Br$\gamma$, but are 
included to indicate the intrinsic CO index of the underlying 
population.

\begin{table}
 \centering
 \begin{minipage}{70mm}
  \caption{CGS4 Slit Parameters for NGC~2903.}
  \begin{tabular}{@{}lccc@{}}
\hline
   Slit    & P.A.          & Clusters & $t_{\rm exp}$ \\
\hline
  1 &    165.6 &   1,4,(nucleus),9,11  &  5760\\
  2 &    148.9 &       3,6,7,16        &  5760\\
  3 &     26.0 &     2,3,(nucleus)     &  5760\\
  4 &     11.9 &      1,2,3,17         &  4560\\
  5 &     90.0 &         none         &  9000\\
\hline
\end{tabular}

Notes.--- See Table~3 for identification of continuum
knots. Cluster number 16 is near the edge of $H$-band image and 
cluster 17 is too faint, and are not included in the photometry 
of Table~3.
\end{minipage}
\end{table}

\begin{table}
 \centering
 \begin{minipage}{70mm}
  \caption{Spectroscopy of the nucleus and bright stellar
clusters in NGC~2903.}
  \begin{tabular}{@{}lccc@{}}
\hline
Cluster(s) &  ${\rm EW(Br}\gamma$)  & ${\rm CO}_{\rm sp}$ &         
${\rm CO}_{\rm ph}$\\
    & (\AA) \\
\hline
                            Slit 1\\
\hline
1+9  & $2.4 \pm 0.2$ & $0.27 \pm 0.01$ & $0.20 \pm 0.01$\\
nucleus  & $2.3 \pm 0.8$ & $0.22 \pm 0.01$ & $0.16 \pm 0.01$\\
4+11 & $8.7 \pm 0.4$ & $0.21 \pm 0.01$ & $0.16 \pm 0.01$\\
\hline
                            Slit 2\\
\hline
6    & $4.6 \pm 0.3$ & $0.25 \pm 0.01$ & $0.18 \pm 0.01$\\
3    & $2.1 \pm 0.7$ & $0.32 \pm 0.01$ & $0.23 \pm 0.01$\\
7    & $3.5 \pm 0.5$ & $0.28 \pm 0.01$ & $0.21 \pm 0.01$\\
17   & $<1.9$         & $0.26 \pm 0.02$ & $0.19 \pm 0.02$\\
\hline
                            Slit 3\\
\hline
nucleus  & $2.5 \pm 0.2$ & $0.22 \pm 0.01$ & $0.16 \pm 0.01$\\
2+3  & $4.7 \pm 0.3$ & $0.30 \pm 0.01$ & $0.22 \pm 0.01$\\
\hline
                            Slit 4\\
\hline
1+9  & $3.8 \pm 0.2$ & $0.27 \pm 0.01$ & $0.20 \pm 0.01$\\
2+3  & $4.7 \pm 0.2$ & $0.30 \pm 0.01$ & $0.22 \pm 0.01$\\
18   & $1.7 \pm 0.3$ & $0.26 \pm 0.02$ & $0.19 \pm 0.02$\\
\hline
                            Slit 5\\
\hline
E    & $<1.1$         & $0.21 \pm 0.02$ & $0.16 \pm 0.02$\\
W    & $<3.2$         & $0.20 \pm 0.02$ & $0.15 \pm 0.02$\\
\hline
\end{tabular}

\end{minipage}
\end{table}

\section{Morphology}

Figure~1 presents the CFHT/PUEO AO $K^\prime$-band image and the 
{\it HST}/NICMOS continuum-subtracted 
Pa$\alpha$ image of the central 46\,arcsec \ 
($\simeq 2\,$kpc) of NGC~2903. 
The continuum light (see also Figure~2, left panel 
for a close-up of the $H$-band emission) resolves 
the already known infrared hot spots (e.g., Wynn-Williams \& Becklin
1985; Simons et al. 1988; P\'erez-Ram\'{\i}rez et al. 2000) into a 
large number of individual stellar clusters. 

\begin{figure}
\psfig{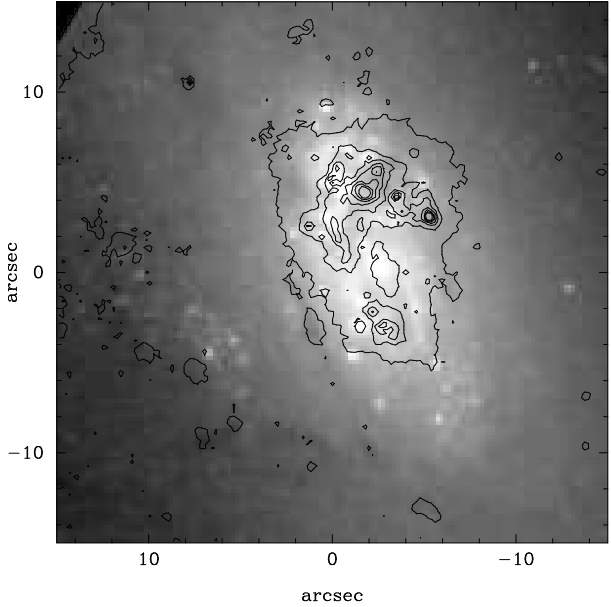}

\caption{$H$-band (NIC3 F160W) continuum image (grey scale) on a logarithmic
scale and, superimposed, the continuum subtracted Pa$\alpha$ 
image (black contours on a linear scale).}
\end{figure}

The most striking feature of the emission in NGC~2903 is the lack 
of spatial correspondence between the bright H\,{\sc ii}
regions seen in the Pa$\alpha$ image and the position of the 
stellar clusters (see superposition of the 
Pa$\alpha$ emission on the $H$-band emission in 
Figure~5). Previous works of this galaxy at lower
spatial resolution identified the bright infrared hot spots 
with sites of intense SF (e.g., Wynn-Williams \& Becklin 1985). It now
appears that the bright NIR sources although close to the 
H\,{\sc ii} regions,  are not coincident with them. 
Other high resolution NIR studies of star forming
galaxies have found a similar result and interpret 
the bright infrared stellar clusters as 
the result of the evolution of giant H\,{\sc ii} regions 
(see for instance, Alonso-Herrero et al. 2000, 2001). The 
properties of the stellar clusters and the H\,{\sc ii} regions 
in NGC~2903 are discussed in Sections~5 and 6.

The {\it HST}/NICMOS Pa$\alpha$ line emission map reveals, for the first time, 
the presence of a nuclear ring-like morphology with an apparent 
diameter of approximately $15\,{\rm arcsec} =625\,$pc. 
The ring is composed 
of a number of bright individual H\,{\sc ii} regions embedded in a 
diffuse component. Thanks to the reduced extinction in the NIR,  
this ring shows up more prominently in Pa$\alpha$  than in 
H$\alpha$ observations of the
nuclear region of NGC~2903 (see for instance Planesas, Colina, 
\& P\'erez-Olea 1997). The southern part of the ring is 
fainter than the northern part; the $H-K^\prime$ colour map shows
that this is not an extinction artifact. 

The right panel of Figure~2
shows the $m_{\rm F160W}-K^\prime$ (equivalent to a ground-based
$H-K^\prime$) colour map constructed with the $H$-band 
{\it HST}/NICMOS
image and the AO $K^\prime$-band images. Colour maps with a 
larger field of view can be found in 
P\'erez-Ram\'{\i}rez et al. (2000). The $H-K^\prime$ colour map (which is 
assumed to trace mainly the extinction to the stars) reveals 
the complex morphology of the obscuration in the centre of this galaxy. The 
bright stellar clusters may appear artificially bluer due 
to the different point spread functions of the AO and NICMOS
images.  The extinction appears to be higher in regions to the north of 
the nucleus, where the Pa$\alpha$ emission is also brighter, ruling 
out the possibility that extinction is hiding the Pa$\alpha$ emission 
from the southern part of the ring of SF. 
This asymmetric distribution of the dust in the central regions of
NGC~2903 was already noticed from lower spatial resolution 
studies involving optical and NIR imaging (Simons et 
al. 1988). 

The asymmetry seen in the H\,{\sc ii} region and 
obscuration distributions is also
apparent in the distribution of the bright stellar clusters, whose
number is more elevated to the north of the centre of NGC~2903.
On large scales, both the gas and continuum emissions
appear  very symmetric (see e.g. the $K$-band 
image in Regan \& Elmegreen 1997; the H$\alpha$ image in Jackson et al.
1991; and the optical broad-band and H$\alpha$ images in P\'erez-Ram\'{\i}rez 
\& Knapen 2000). 

\begin{figure}
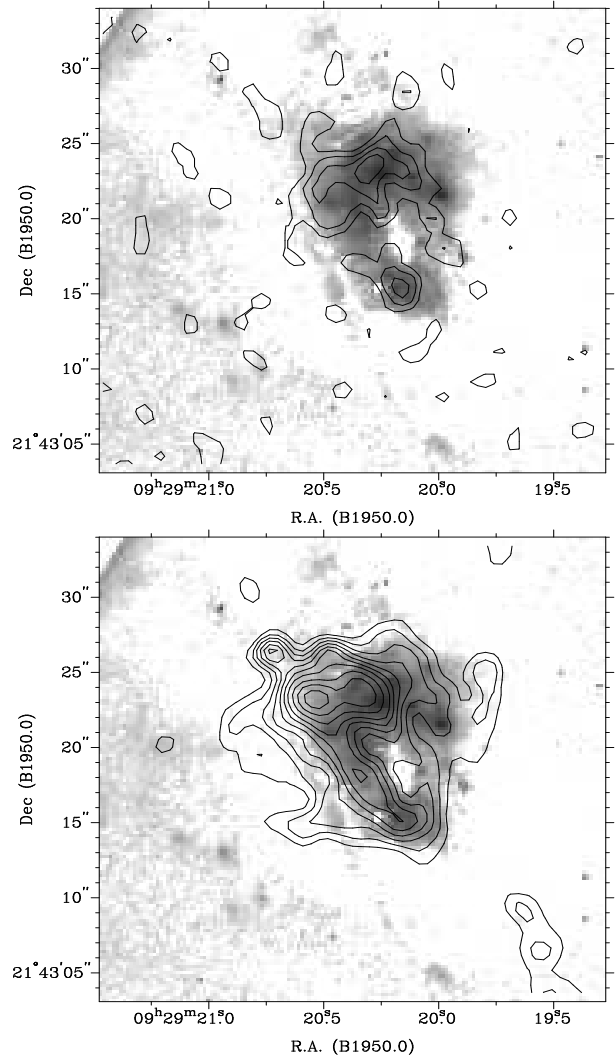

\psfig{file=figure6a.ps,height=7cm,width=8cm,rheight=7cm,rwidth=0cm,angle=-90}
\psfig{file=figure6b.ps,height=7cm,width=8cm,rheight=7cm,rwidth=0cm,angle=-90}

\caption{{\it Upper panel:} 
Continuum subtracted Pa$\alpha$ (grey scale) on a logarithmic 
scale and 2\,cm radio map (black contours) on a linear scale 
from Wynn-Williams \& Becklin (1985). {\it Lower panel:} as in upper 
panel but the contours are the 20\,cm radio map. The resolution 
of the radio maps is approximately 2\,arcsec (FWHM).}
\end{figure}

Another puzzling result from previous NIR studies of this 
galaxy was the lack of spatial correspondence between the 
optical and $2.2\,\mu$m continuum 
morphologies and the radio emission (Wynn-Williams
\& Becklin 1985; Simons et al. 1988). However when we 
superimpose the radio emission maps (at 2\,cm and 20\,cm) 
from Wynn-Williams \&
Becklin (1985) on the {\it HST}/NICMOS Pa$\alpha$ image for which 
we used the astrometry information as given in the headers of the image 
data files) we find an 
excellent overall correspondence between the line emission 
and the radio morphologies (Figure~6), as did Planesas et al. 
(1997). The north and south peaks detected 
in the 2\,cm radio image (Wynn-Williams \&
Becklin 1985) are clearly identified with H\,{\sc ii}
region emission in the ring of SF. This is in 
good agreement with the result that a significant fraction of the 
radio emission in this galaxy is  thermal in origin 
(free-free emission from H\,{\sc ii} regions, Wynn-Williams 
\& Becklin 1985). Additional support comes 
from the radio spectral indices of the north and south peaks detected
in the radio maps (Wynn-Williams \& Becklin 1985), and from the 
good agreement between the predicted free-free emission at 2\,cm 
from the number of ionizing photons (Simons et al. 1988) and the 
observed radio flux densities. The lack of a detailed correspondence 
at the level of the individual emitting regions is probably due
to resolution effects, since the resolution of the radio maps
is an order of magnitude poorer than that of the {\it HST}/NICMOS 
Pa$\alpha$ map.

\section{Extinction}

\begin{table*}
 \centering
\begin{minipage}{110mm}
  \caption{Photometry of the nucleus of NGC~2903 and bright stellar clusters.}
  \begin{tabular}{@{}lcccccc@{}}
\hline
   Source     & x off           & y off & $m$(F160W) &
$K^\prime$  & $m_{\rm F160W}-K^\prime$ & $M_{\rm F160W}$ \\
              & (arcsec) & (arcsec) \\[10pt]
\hline
Nucleus  & 0.     &0.        &12.99  &12.55   &0.44 & $-16.68$\\
 1       & $+0.9$ & $-3.4$ &14.64  &14.46   &0.18 & $-15.03$\\
 2       & $+2.6$ &+4.8     &15.04  &14.61   &0.43 & $-14.62$\\
 3       & $+2.0$ &+3.9     &15.75  &15.45   &0.30 & $-13.92$\\
 4       & $-1.1$ &+4.0     &15.76  &15.54   &0.22 & $-13.92$\\
 5       & $+1.8$ &+4.3     &15.77  &15.41   &0.36 & $-13.90$\\
 6       & $+2.9$ &+2.6     &15.77  &15.55   &0.22 & $-13.90$\\
 7       & $+0.6$ &+6.2     &15.82  &15.65   &0.17 & $-13.85$\\
 8       & $+0.8$ &+1.0     &15.90  &15.71   &0.19 & $-13.75$\\
 9       & $+1.1$ &$-5.2$  &16.00  &15.87   &0.13 & $-13.67$\\
 10      & $+2.3$ &+5.4    &16.26  &15.26   &1.00 & $-13.41$\\
 11      & $-0.8$  & +3.0    &16.46  &16.41   &0.05 & $-13.21$\\
 12      & $-2.3$  &$-2.6$  &16.48  &15.97   &0.51 & $-13.19$\\
 13      & $-3.5$    &$-5.6$  &16.52  &16.33   &0.19 & $-13.15$\\
 14      & $+1.7$ & +6.5    &16.57  &16.49   &0.08 & $-13.10$\\
 15      & $+3.1$ & +1.7    &16.63  &16.19   &0.44 & $-13.04$\\[10pt]
\hline
\end{tabular}

Notes.--- The photometry 
of the nucleus is through a 1.4\,arcsec-diameter 
aperture, whereas the photometry of the stellar clusters 
is through a 0.8\,arcsec-diameter aperture. 
The absolute F160W ($H$-band) magnitudes are not corrected for extinction.

\end{minipage}
\end{table*}

In Figure~2 (right panel) 
we show an $H-K^\prime$ colour map which traces the extinction 
to the stars. An advantage of using a NIR colour map to 
derive the extinction is that the NIR colours do not
depend on  the age of the stellar population as strongly as 
optical colours (see Section~5.3). Assuming a colour of the unreddened 
stellar population of $H-K^\prime =0.2$ and a simple foreground 
dust screen model we find values of the visual extinction 
to the stars of up to $A_V=9.5\,$mag using Rieke \& Lebofsky's (1985)
extinction law. The average value of the extinction over the field
of view of the colour map is  
$A_V = 2.3\,$mag, equivalent to an extinction in the $H$-band of
$A_H = 0.4\,$mag. In Table~3 we give the $m_{\rm F160W} - K^\prime$
colour (equivalent to a ground-based
$H-K^\prime$ colour) of the nucleus of NGC~2903. In a similar 
fashion as above we derived 
an extinction to the nucleus of NGC~2903 of $A_V =  
3.8\,$mag, which is relatively modest compared with the higher 
values found to the north of the nucleus where most of 
the SF is occurring.

Beck et al. (1984) derived a value of the visual extinction to the
gas within the central 8\,arcsec \ of $A_V=7\,$mag from the
NIR hydrogen recombination lines Br$\gamma$ and Br$\alpha$. We 
made use of these line measurements and our 
Pa$\alpha$ line flux measured through the 
same aperture, and Rieke \& Lebofsky (1985)
extinction law. The intrinsic values of the 
line ratios (for case B recombination) 
were taken from Hummer \& Storey (1987) 
for the assumed physical conditions of the gas 
$T_{\rm e} = 10^4\,$K and $n_{\rm e} = 100\,{\rm cm}^{-2}$. 
We derive values for the visual extinction of 
$A_V = 7.0\pm2.5\,$mag from Pa$\alpha$/Br$\gamma$, 
$A_V = 5.8\pm0.9\,$mag from Pa$\alpha$/Br$\alpha$, 
and $A_V = 5.1\pm1.5\,$mag from Br$\alpha$/Br$\gamma$.
The errors in the extinction take into account a 10\%
uncertainty in the Pa$\alpha$ flux calibration. All the 
values of the extinction to the gas  are 
consistent to within the observational errors, and 
are in relatively good agreement with the average value of the extinction 
to the stars over the same region from the $H-K^\prime$ colour map
($A_V \simeq 3\,$mag).

\begin{figure}
\hspace{-2cm}
\psfig{file=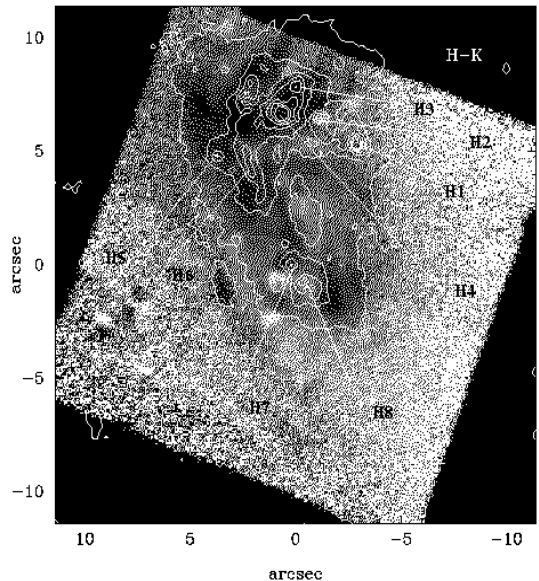,height=10cm,width=12cm,rheight=10cm,rwidth=0cm,angle=0}
\vspace{-1cm}
\caption{$H-K^\prime$ colour map (grey scale, as in Figure~2)
with the continuum subtracted Pa$\alpha$ image (white contours) 
superimposed on a linear scale. The names of the brightest 
H\,{\sc ii} regions are indicated (see Table~4).}
\end{figure}

\begin{table*}
 \centering
 \begin{minipage}{130mm}
  \caption{Photometry of the bright H\,{\sc ii} regions 
in the ring of SF.}
  \begin{tabular}{@{}lccccccc@{}}
\hline
   Source     & x off           & y off & $f({\rm Pa}\alpha)$ &
$A_V$ & $\log L ({\rm H}\alpha)$ & $\log N_{\rm Ly}$ &
optical \\
              & (arcsec) & (arcsec) & (erg cm$^{-2}$ s$^{-1}$) 
& (mag) & (erg s$^{-1}$) & (${\rm s}^{-1}$) & counterpart\\[10pt]
\hline
H1 & $-2.8$ & +2.8 & $2.06 \times 10^{-14}$ & 
$1.8\pm0.4$ & 39.31 & 51.18 & R2 \\
H2 & $-1.0$ & +3.8 & $1.49 \times 10^{-14}$ & 
$1.8\pm0.4$ & 39.17 & 51.03 & R1 \\
H3 & 0.0  & +5.4 & $2.13 \times 10^{-14}$ & 
$2.7\pm0.4$ & 39.37 & 51.23 & R8 \\
H4 & +0.7 & +4.1 & $3.32 \times 10^{-14}$ & 
$3.8\pm0.4$ & 39.63 & 51.50 & R7 \\
H5 & +3.6 & +2.2 & $1.04 \times 10^{-14}$ & 
-- & 38.90 & 50.77 & --\\
H6 & +2.4 & +2.4 & $1.62 \times 10^{-14}$ & 
$1.8\pm0.4$ & 39.20 &51.07& R6 \\
H7 & +0.4 & $-2.6$ & $0.94 \times 10^{-14}$ & 
 -- & 38.86 & 50.72 & -- \\
H8 & $-0.4$ & $-3.2$ & $1.31 \times 10^{-14}$ & 
$2.2\pm0.4$ & 39.13 & 51.00 & R4 \\[10pt]
ring & -- & -- & $5.28\times10^{-13}$ & 
$5.8\pm 0.9$ & 40.95  & 52.82 &  --\\[10pt]  
\hline
\end{tabular}

Notes.--- Columns~(2) and (3) Offsets with respect to the 
nucleus of NGC~2903 (peak of continuum emission). Column~(4) Observed
Pa$\alpha$ flux through a 1.2''-diameter aperture (50\,pc). 
Column~(5) Visual extinction to the gas as derived 
from the observed Pa$\alpha$ to H$\alpha$ line ratios using
a foreground dust screen model. 
Column~(6) H$\alpha$ luminosities (corrected for extinction
when available). Column~(7) Number of ionizing photons. 
Column~(8) Correspondence with the 
H$\alpha$ sources detected in Planesas et al. (1997).
The total ring Pa$\alpha$ flux and H$\alpha$ luminosity 
(the latter corrected for extinction) correspond to 
a 14\,arcsec-diameter aperture.
\end{minipage}
\end{table*}

We can carry out a more detailed study of the spatial 
distribution of the extinction to the 
gas at the location of the individual H\,{\sc ii} regions using our 
Pa$\alpha$ image and the H$\alpha$ fluxes 
given by Planesas et al. (1997). We have identified 
8 bright H\,{\sc ii} regions in the Pa$\alpha$ image 
(see Figure~7 and Table~4). A careful cross-correlation 
between the positions of these H\,{\sc ii} regions 
seen in the Pa$\alpha$ image (see Table~4 for 
relative positions with respect to the peak of
continuum emission) and 
those identified by Planesas et al. (1997) from their
H$\alpha$ image shows that there are six correspondences 
(see Table~4). We used 
the 2\,arcsec-diameter (circular aperture) H$\alpha$ fluxes 
from Planesas et al. 
(1997) and Pa$\alpha$ line flux measurements for the H\,{\sc ii}
regions seen in both images (thus, relatively low extinction
regions). Again a simple foreground dust 
screen model is assumed for the geometry of the gas. 
We list the values of the visual 
extinction and the corresponding errors in Table~4; these are 
between $A_V =1.8\,$mag and $A_V= 3.8\,$mag using a
foreground dust screen model.
 The errors of the extinction were computed assuming 
a 10\% uncertainty in the Pa$\alpha$ fluxes. The 
fact that regions H5 and H7 are not seen in the H$\alpha$ 
image is probably indicating that the extinction to these 
regions is elevated. The foreground dust screen model always
gives a lower limit to the extinction. In a more realistic
model with a uniform distribution of stars mixed homogeneously 
with dust, the derived values of the visual extinctions are 
between $A_V=5$ and $A_V=15\,$mag for the sources in Table~4
(except for source H4).

The values of the extinction 
to the gas of the H\,{\sc ii} regions are slightly smaller
than the corresponding values of the extinction to the 
stars at the same positions as derived from the 
$H-K^\prime$ colour map, although the relative values seem to 
be in good agreement (when the foreground dust
screen model is used). For instance from Figure~7 
it is clear that H4 lies in a region  with higher 
extinction than H1 or H2. Regions H5 and H7 (which are not
detected in the optical) are faint and are located close
to regions of high extinction.  
The discrepancies between the extinction 
values obtained from the $H-K^\prime$ colour map and the hydrogen 
recombination lines involving H$\alpha$ are most likely due 
to the simplistic assumption for the dust geometry as well as to optical 
depth effects in the regions with higher extinction. Using 
H$\alpha$ fluxes will bias the extinction calculations towards 
the outer regions which are dominated by 
lower obscuration (see discussion in Alonso-Herrero
et al. 2000). Summarizing, although the distribution of the 
obscuring material in the central regions of NGC~2903 is quite
complex, the values found for the extinction in 
the NIR are relatively modest, and therefore for the 
analysis in the next two sections we will assume that the 
extinction does not play a relevant role.

\section{The young stellar clusters}

Recent high resolution {\it HST} and ground-based studies have 
shown growing evidence for a conspicuous presence of young stellar clusters 
in a variety of galactic environments: interacting galaxies 
(the Antennae galaxy, Whitmore et al. 1999 and references therein), 
galaxies with CNR (Barth et al. 1995; Ryder \& Knapen 1999), 
infrared luminous and ultraluminous galaxies (Surace et al. 1998; 
Alonso-Herrero et al. 2000, 2001; Scoville et al. 2000). 

The properties, nature and 
evolution of these young stellar clusters are relevant for the 
understanding of the processes triggering the SF 
in such diverse environments. In what follows we will study in 
detail the properties of the stellar clusters in NGC~2903 making
use of both the spectroscopic and imaging information.

\subsection{Luminosities}

We used {\it daophot} within {\sc iraf} to identify stellar clusters
on the NIC2 F160W image (which is the image with the highest 
angular resolution). We found
71 individual clusters in the central 20\,arcsec \ (830\,pc) of 
NGC~2903. From the larger field of view  of the NICMOS (NIC3) and AO images, 
it is apparent that most of the bright stellar clusters are confined within 
the nuclear region, and thus they are representative of 
the population of bright stellar clusters. 
We performed aperture photometry through a 0.61\,arcsec-diameter 
(corrected for aperture effects, M. Rieke 1999, private communication). 
The surrounding (underlying) galaxy emission was subtracted 
using the mean background value from within 
an annulus from 1.0\,arcsec \ and 1.4\,arcsec \  around each source. 

\begin{figure}
\psfig{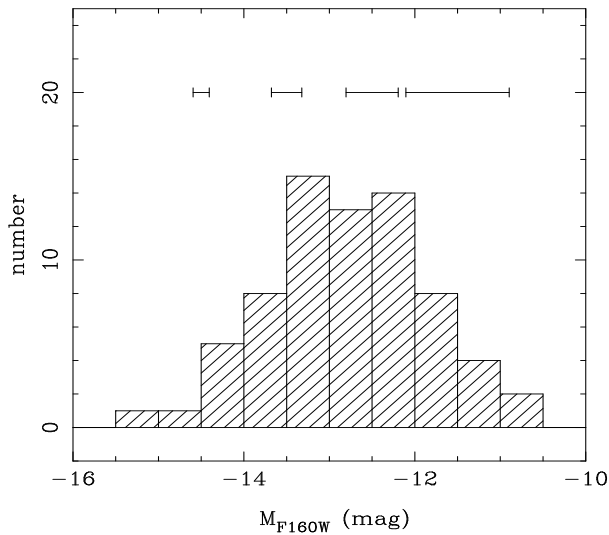}
\caption{Histogram of the absolute $H$-band (NIC2 F160W) 
magnitudes of the brightest stellar clusters in the central 
region of NGC~2903. The errors are average values per magnitude 
interval.}
\end{figure}

Figure~8 shows a histogram of the absolute F160W magnitudes (equivalent to
ground based absolute $H$-band magnitudes) for
the brightest 71 clusters in the central region of 
NGC~2903, using a distance modulus of $m-M = 29.67$. The 
absolute magnitudes have not been corrected 
for extinction. The effect of this correction would 
be small since the average 
extinction in the $H$-band is relatively 
small ($A_H=0.4\,$mag) if the foreground dust screen model is
used. The average absolute magnitude is  $M_H = 
-13\,$mag. The errors per magnitude 
interval were computed assuming a typical error of 10\%
associated with the galaxy background subtraction. 
In Figure~8 we show the average value of the error
per magnitude interval. 
The actual value for a given cluster 
depends on the brightnesses of both the source 
and the local background. Our cluster detection is not complete, 
and we anticipate that most of the faint stellar clusters have been 
missed. The detection efficiency varies with the brightness of the source, 
local galaxy background and crowding. 

The luminosities of the 
stellar clusters detected in NGC~2903 are similar to those in 
M100 (Ryder \& Knapen 1999) or NGC~1097 (Barth et al. 1995), both 
are barred galaxies which contain bright hot spots as 
in NGC~2903. Moreover, the luminosities of the stellar clusters 
in NGC~2903 are comparable to those of the brightest (young)
stellar clusters found in the interacting galaxy The 
Antennae (Whitmore et al. 1999) which range from $M_H =-13.3\,$mag up to
$M_H = -15.9\,$mag (assuming a typical colour of $V - H \simeq 2$), although
fainter than those detected in the infrared interacting galaxies 
Arp~299 and NGC~1614 (Alonso-Herrero et al. 2000, 2001). 

Some of the clusters are unresolved, but the
typical size of 8\,pc (FWHM)  is  in good agreement 
with the sizes measured for  the stellar clusters in the Antennae
galaxy ($R_{\rm eff}  = 0-10\,$pc, Whitmore et al. 1999).

\subsection{Ages}
The spectroscopic information of the bright stellar clusters (EW 
of Br$\gamma$ and the values of the CO index) is a valuable tool to
constrain their ages. The EWs of hydrogen recombination
lines are a good measure of the age of the starburst since the
luminosity of the lines is directly proportional to the number of ionizing
photons. In particular, the EW of Br$\gamma$ measures the number 
of young stars (producing the ionizing photons) relative to the 
number of red supergiants. The strength of the $2.29\,\mu$m CO band is
sensitive to the stellar effective temperature and 
surface gravity and it is often used to
determine the existence of red supergiants. Thus a diagram of 
the CO index versus the EW of Br$\gamma$  can be used  
to constrain in detail the ages of
the SF episodes (e.g., Puxley et al. 1997; Vanzi, Alonso-Herrero,
\& Rieke 1998; Ryder \& Knapen 1999).

\begin{figure}
\psfig{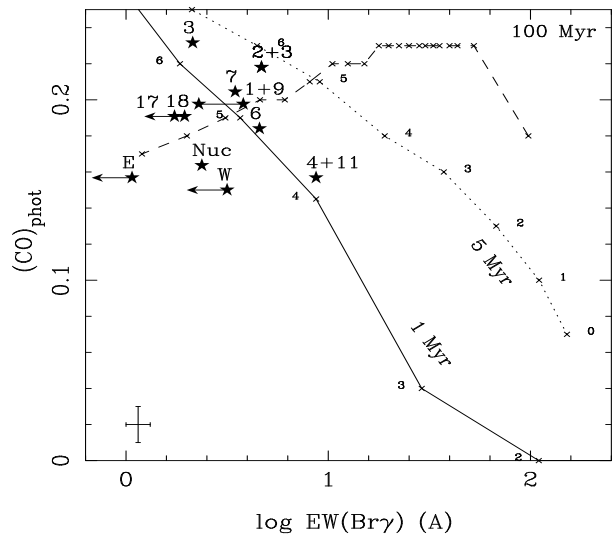}

\caption{Photometric CO index versus EW of 
Br$\gamma$. The lines are models (Rieke et al. 1993) 
with a Gaussian burst with the FWHM indicated next to the curves. 
The crosses on the 1\,Myr and 5\,Myr curves are the ages after
the peak of SF indicated by the small numbers next to them. 
The filled stars are the measurements for the 
stellar clusters (Table~2).
The typical error bars of our measurements are plotted in the bottom
left corner. The arrows indicate an upper limit on the EW of 
Br$\gamma$. The two values of the EW of Br$\gamma$ for 1+9 
(Table~3) are connected with a line.}
\end{figure}

We use Rieke et al. (1993) models with a single burst of SF 
with Gaussian FWHM values of 1\,Myr, 5\,Myr
and 100\,Myr. The longest Gaussian burst
represents a slowly evolving population with 
quasi constant SF. For the IMF slope we chose a 
Salpeter form with lower and upper mass cutoffs of 1 and $80\,{\rm M}_\odot$
respectively. 
The outputs from the models together with the observations of the 
bright stellar clusters are shown in Figure~9. The numbers on 
the 1\,Myr and 5\,Myr curves represent the age after the 
peak of SF which occurs at $\simeq 2.5\times 10^{6}\,$yr
and $\simeq 7 \times 10^6\,$yr after the onset of
the SF, respectively. For the 100\,Myr curve the crosses are in 
10\,Myr intervals and the first point to the right indicates 
100\,Myr before the peak of SF.   

The interpretation of this figure to derive the ages of the 
clusters may be somewhat
problematic when only a few points are available, because the observations
are sometimes located close to several curves. Moreover, 
for the form of the SF chosen here the 100\,Myr curve
crosses both the 1\,Myr and 5\,Myr curves, and a degeneracy 
between the total duration of the episode of SF and 
its age may occur. In the case of the bright stellar 
clusters in NGC~2903, however,
it is clearly seen from the figure that most of the points are 
well reproduced by the models, and 
in fact, they lie parallel to the 1\,Myr and 5\,Myr model outputs. This 
suggests that short-duration bursts are responsible for the 
properties of the stellar clusters, rather than constant star 
formation. The ages measured from the peak 
of the SF of most of the clusters show a very narrow 
range, between $4 \times 10^6\,$yr and $7 \times 10^6\,$yr.  The 
observed values for 
the E and W positions indicate an older population, in good
agreement with their location outside  the SF ring. 
The position of the nucleus on this diagram is ambiguous, although it
shows values similar to  those of the E and W positions.
The absolute ages of the knots 
depend upon the specific form assumed for the SF 
with ranges $10-14 \times 10^6\,$yr and $6.5-9.5\times 10^6\,$yr 
for the 1\,Myr and the 5\,Myr models respectively.

\subsection{Masses}
We can also obtain an estimate of the masses of the stellar clusters 
using evolutionary synthesis models, now in combination with a  
NIR colour-magnitude diagram. An advantage 
of using a NIR colour-magnitude diagram is that the effects
of extinction are greatly reduced as compared with the optical. 
In Table~3 we list the $m_{\rm F160W}$ and $K^\prime$ 
aperture photometry, colours and F160W absolute magnitudes  
for the fifteen brightest clusters. Typical errors in 
the colours are  $\simeq 0.20\,$mag, and are dominated by the 
uncertainties due to  the galaxy background subtraction. 

\begin{figure}
\psfig{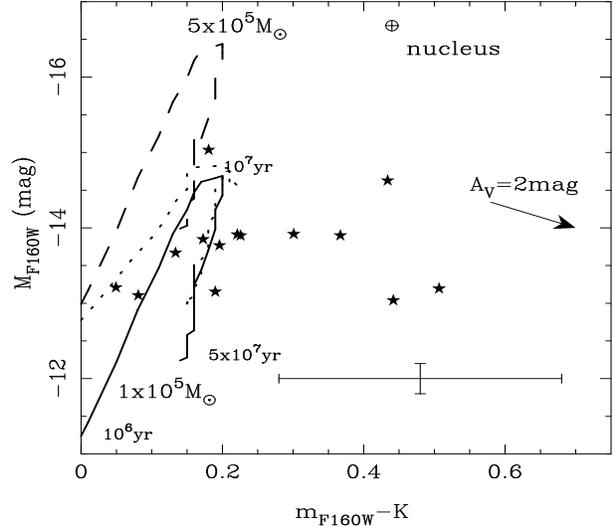}

\caption{Colour-magnitude diagram for the brightest 15 stellar clusters
(filled stars, Table~3). 
We also plot outputs from Rieke et al. (1993) starburst models 
with Gaussian bursts of SF with ${\rm FWHM} = 1\,$Myr  
normalized to a mass of $10^5\,{\rm M}_\odot$ (dotted line), and 
models with ${\rm FWHM} = 5\,$Myr normalized to a $10^5\,{\rm M}_\odot$ 
mass (solid line) and to a $5 \times 10^5\,{\rm M}_\odot$ mass 
(dashed line). We use a Salpeter IMF.   
The ages shown next to the solid line model show the evolution from 
$10^6\,$yr (lower left part of the curve) 
up to $50 \times 10^6\,$yr after the peak of SF (lower right part of 
the curve), with the peak of the $M_{\rm F160W}$ magnitude at
around $8 \times 10^6\,$yr.}
\end{figure}

Figure~10 shows the colour-magnitude diagram, where we have plotted 
the nucleus of NGC~2903 and the fifteen brightest stellar clusters
(Table~3). The lines in this diagram are the model outputs from the 
1\,Myr Gaussian burst normalized to a total mass of $10^5\,{\rm M}_\odot$
(dotted line), and the 5\,Myr Gaussian burst normalized to 
a total mass of $10^5\,{\rm M}_\odot$ (solid line) and of 
$5 \times 10^5\,{\rm M}_\odot$
(dashed line). The time evolution of the models 
is shown from 1 (lower left part of the 
curve) up to $50\times 10^6\,$yr (lower right part of
the curve). The peak of the $H$-band absolute magnitude 
occurs $\simeq 6.5\times 10^6\,$yr and 
$\simeq 8\times 10^6\,$yr after the 
peak of SF for the  1\,Myr and 5\,Myr models, 
respectively. The parameters used for the models are similar to those 
used to fit the SF properties of Arp~299 and 
NGC~1614 (see Alonso-Herrero et al. 2000, 2001 for more details).
The arrow shows the effects of two magnitudes of visual extinction.
Note that for most of the clusters a value of $A_V \le 8\,$mag 
(or similarly in the $H$-band $A_H \le 1.4\,$mag) would move the 
observed points to the loci of the evolutionary synthesis models. These
values of the extinction are consistent with the estimates 
using the foreground dust screen model in Section~4.

From this figure it is readily seen 
that all the brightest stellar clusters in NGC~2903  have masses of between 
$10^5\,{\rm M}_\odot$ and $5 \times 10^5\,{\rm M}_\odot$, 
in good agreement with those of 
Galactic globular clusters (van der Bergh 1995).  
As noted  in Section~3, the extinction estimates 
obtained from recombination lines and the mixed
model are higher than for the foreground dust screen model, which would
imply higher masses for the clusters. But because of the lack of spatial 
correspondence between the stellar clusters and the H\,{\sc ii}
regions we cannot be sure that the high values of the extinction 
are correct for the stellar clusters. 
Nevertheless, even if the upper limits to the extinction 
are used, the masses of the stellar clusters in 
NGC~2903 are below those  of the stellar clusters  
found in luminous and ultraluminous galaxies (e.g., Alonso-Herrero et al. 
2000 and references therein). 

We can  quantify the contribution of the young stars in the 
stellar clusters to the total mass in the central region of 
this galaxy from a statistical point of view. In Section~5.2
we have shown that the ages of the brightest stellar clusters 
show a very narrow range. Moreover there is no correspondence between the 
brightness of a knot and its age (Figure~9 and Table~3), that is, the 
brightest sources do not tend to be the youngest (e.g., knot 3 or knots
2+3). Thus we can assume that the brightnesses of the stellar clusters
are proportional to their masses, whereas their age is a second order 
effect. The evolutionary synthesis models provide the evolution of the   
$H$-band mass-to-light ratio and can be used to 
derived the masses of the stellar clusters. Evidently, 
this ratio depends upon the age of the cluster and the form assumed
for the SF (and metallicity). From Figure~10 it is clear
that this NIR mass-to-light ratio will have a minimum at 
an age of around $10^7\,$yr (that is, at the peak 
of the $H$-band luminosity). For the range of ages derived for the stellar 
clusters in NGC~2903 the mass-to-light ratio is uncertain by a 50\%. 
As an illustration of the changes of the infrared mass-to-light 
ratio as a function of the IMF slope and mass cutoffs, type of star 
formation and metallicity 
we refer the reader to figures 51 and 52 in Leitherer et al. (1999) 
models.

The contribution of the young stellar clusters to the total 
$H$-band luminosity can be obtained by adding the luminosities from 
the histogram in Figure~8 ($\simeq 13\%$). Now, using the model 
mass-to-light ratios we find a mass in the young stellar clusters of 
$M_{\rm cluster\,*} = 2.1-3.6 \times 10^8\,{\rm M}_\odot$
within the central $\simeq 625\,$pc of NGC~2903. The range in masses 
takes into account the spread in ages (and thus a varying 
mass-to-light ratio) of the stellar clusters. 
For the old stellar population we use the mass-to-light ratio 
found by Thronson \& Greenhouse (1988) and estimate a mass in old stars of
$M_{\rm old\,*} = 2.7 \times 10^9\,{\rm M}_\odot$ within
the same region. The 
young stellar clusters therefore account for $7-12\%$ of the 
central ($625\,$pc) stellar mass in NGC~2903.  Note, however, that
a population of even younger stars (see next section), which are 
ionizing the gas in the bright H\,{\sc ii} regions, are unaccounted for 
in the mass calculation.

\section{STAR FORMATION PROPERTIES}

\subsection{The star-forming regions}
As shown in Sections~3 and 4, we have identified 
eight bright H\,{\sc ii} regions in the ring of SF
of NGC~2903. These H\,{\sc ii} regions appear 
resolved with typical sizes of between 16\,pc and 50\,pc
(FWHM). Table~4 lists the observed 
Pa$\alpha$ fluxes through a 1.2\,arcsec-diameter  (50\,pc) 
circular aperture, the H$\alpha$ luminosities (corrected for extinction
when possible) and the equivalent number of ionizing photons 
assuming that there are no escaping photons.
The  extinction corrected H$\alpha$ luminosity of the bright 
H\,{\sc ii} region H4 in the ring of SF approaches that of 
the super-giant H\,{\sc ii} region 30 Doradus in 
the LMC ($\log  L({\rm H}\alpha) 
= 39.8\,{\rm erg \ s}^{-1}$, Kennicutt, Edgar,
\& Hodge 1989), but 
it is below the luminosities of the H\,{\sc ii} regions in infrared 
luminous galaxies (Alonso-Herrero et al. 2000, 2001).
However, the bright H\,{\sc ii} regions in the central region of NGC~2903 
are of comparable luminosity as the brightest H\,{\sc ii} 
regions in nearby spiral galaxies (Kennicutt et al.
1989; Rozas et al. 1996; Knapen 1999). 

Although spectroscopic information is not available for the bright
H\,{\sc ii} regions, we can compare the relative values of their 
EW of Pa$\alpha$ with those of the stellar clusters. A map of 
the EW of Pa$\alpha$ can be constructed by dividing the continuum-subtracted 
{\it HST}/NICMOS Pa$\alpha$ line emission image by the $H$-band continuum 
image.
Comparing the relative values of the EW(Pa$\alpha$) at the location of the
H\,{\sc ii} regions H1 and H4 with the youngest stellar cluster 
(4+11), we find that these two H\,{\sc ii} regions have values
of the EW of between three and four times greater than 4+11, indicating ages
for these H\,{\sc ii} regions of less than $3 \times 10^6\,$yr after the 
peak of SF (for the 1\,Myr model). The other H\,{\sc ii} regions in the ring
show values for the EWs similar to that of 4+11, but greater than the 
other stellar clusters.  Again using the outputs from the 
evolutionary synthesis models, we find that 
the number of ionizing photons (Table~4, column 7) of the H\,{\sc ii}
regions implies $\simeq 10^{5}\,{\rm M}_\odot$ of recently formed stars
with ages of $\simeq 1.5-4.5\times 10^6\,$yr after the peak of
star formation using the 1\,Myr model. 
In conclusion, the bright H\,{\sc ii} regions
in the star forming ring are younger than the stellar clusters, but 
with similar stellar masses. This 
strongly suggests that these H\,{\sc ii} regions are the progenitors of the 
stellar clusters.

\subsection{The SF efficiency}

The morphology of the nuclear star forming ring 
revealed by the {\it HST}/NICMOS images allows an accurate estimate of the
SF efficiency in this galaxy. Using the well-known relation 
between the SF rate (SFR) and  the H$\alpha$
luminosity (e.g., Kennicutt 1998), and the 
extinction corrected H$\alpha$ luminosity (Table~4),
we find ${\rm SFR} = 0.7\,{\rm M}_\odot\,{\rm yr}^{-1}$ 
in the nuclear ring of NGC~2903.
This value corresponds to a SFR surface density of 
$\log \Sigma_{\rm SFR} = 0.37\,{\rm M}_\odot\,{\rm yr}^{-1}\,{\rm kpc}^{-2}$. 
The total mass in molecular gas within the central 770\,pc of 
NGC~2903 has been estimated by Jackson et al.
(1991) using CO observations and the standard Galactic CO-to-H$_2$ conversion 
factor to be $3 \times 10^8\,{\rm M}_\odot$ and the molecular gas 
density is  $\log \Sigma_{{\rm H}_2} = 2.60\,{\rm M}_\odot\,{\rm pc}^{-2}$
(Kennicutt 1998). The use of the standard Galactic CO-to-H$_2$ factor is 
uncertain in  general in external galaxies 
(e.g., Adler et al.  1992; Rand 1993; Nakai \& Kuno 1995), 
and certainly so in environments like the one under consideration. 
It is conceivable that SF locally causes an increase in the emission by CO
molecules, as cautiously hinted at by, e.g., Adler, Lo, \& Allen
1991) and Allen (1992). This would
not be accompanied by an increase in the underlying molecular gas mass, as
would be expected through the use of a constant ratio. Given the amount of 
massive SF in the core region of 
NGC~2903, we may expect the CO measurements there in combination with the use 
of a standard ratio value to lead  to an overestimate of the amount of 
molecular hydrogen, and thus to an underestimate of the SF efficiency 
derived from it. 
In any case, the numbers derived above for $\log\Sigma_{\rm SFR}$ and $\log 
\Sigma_{{\rm H}_2}$ would show that the global SF 
efficiency in NGC~2903 is significantly higher than for normal galaxies, 
and it is in fact at the lower end of the relation 
found for infrared luminous star forming
galaxies, which can approach $\simeq 100\%$ efficiencies over 
$10^8\,$yr (Kennicutt 1998). 
\section{Summary and Conclusions}
We have presented  high resolution NIR imaging and spectroscopy of the nuclear
region of NGC~2903 and conducted a detailed study of its morphology 
and SF properties. Our main conclusions are:

\begin{itemize}
\item The bright infrared hot spots (known from ground-based observations) 
are resolved into individual stellar clusters or groups of
these in our AO and {\it HST}/NICMOS images. 

\item The {\it HST}/NICMOS Pa$\alpha$ (at $1.87\,\mu$m)
image reveals a ring of star formation (diameter $\simeq 625\,$pc) with
bright H\,{\sc ii} regions. The overall morphology of the Pa$\alpha$
emission is  similar to the 2\,cm and 20\,cm radio morphologies, 
which seems to indicate a significant 
contribution of thermal emission to the total radio emission.

\item The high resolution images show a complex distribution of the
extinction in the CNR of this galaxy. However the derived values
of the extinction are relatively modest (with values of up to 
$A_V = 9.5\,$mag or infrared extinction at the $H$-band of 
up to $A_H = 1.6\,$mag for a foreground dust screen model).

\item We combined spectroscopic and photometric 
information available for the stellar clusters with evolutionary synthesis 
models to constrain their SF properties.  Short-lived bursts of
SF (Gaussian bursts with ${\rm FWHM} = 1, \, 5\,$Myr) produced
a good fit to the data. The stellar clusters have luminosities 
similar to the young stellar 
clusters in the Antennae galaxy. They are relatively young, with  a 
narrow range of ages, $4-7\times 10^6\,$yr after 
the peak of SF  (the absolute ages depend upon the specific
duration of the burst of SF), and 
contribute to some $7-12\%$ of the stellar mass in the central 
regions of NGC~2903.

\item The H\,{\sc ii} regions in the star forming ring exhibit
luminosities approaching that of 30 Doradus, and are younger
($1-4 \times 10^6\,$yr after the peak of SF) than the stellar clusters.

\item The ages and masses derived for the bright stellar clusters 
and H\,{\sc ii} regions suggest that the H\,{\sc ii} regions are  
the progenitors of the bright infrared clusters. This may 
offer an explanation for the lack of a detailed spatial correspondence 
between the clusters and the H\,{\sc ii} regions.

\item Finally, we show that the SF efficiency in 
the central kpc of NGC~2903 is more
elevated than in normal galaxies, similar to (although 
slightly smaller than) those for infrared luminous galaxies. 
 
\end{itemize}

\section*{Acknowledgments}

This work is partly based on observations collected at UKIRT. The UKIRT
is operated by the Joint Astronomy Centre on behalf of the 
UK Particle Physics and Astronomy Research Council. Based
on observations with the NASA/ESA Hubble
Space Telescope, obtained at the Space Telescope Science Institute, which
is operated by the Association of Universities for Research in Astronomy,
Inc. under NASA contract No. NAS5-26555.

We are grateful to Dr. C. G. Wynn-Williams for providing us with the radio
data.

We thank Drs. Ren\'e Doyon, Dolores P\'erez-Ram\'{\i}rez, and 
Daniel Nadeau for help with the CFHT observations, and Dr. 
Marianne Takamiya for help with those on UKIRT.

\label{lastpage}

\newpage

\normalsize
\end{document}